\documentclass[aps,twocolumn,pra,superscriptaddress,showpacs,tightenlines]{revtex4-1}
\usepackage{multirow}
\usepackage{array}
\usepackage{amsmath}
\usepackage{graphicx}
\usepackage{color}
\usepackage{amsfonts}
\usepackage{txfonts}
\usepackage[colorlinks,citecolor=blue]{hyperref}
\begin{document}

\title{Controllable generation of mechanical quadrature squeezing via dark-mode engineering in cavity optomechanics}

\author{Jian Huang}
\affiliation{Key Laboratory of Low-Dimensional Quantum Structures and Quantum Control of Ministry of Education, Key Laboratory for Matter Microstructure and Function of Hunan Province, Department of Physics and Synergetic Innovation Center for Quantum Effects and Applications, Hunan Normal University, Changsha 410081, China}

\author{Deng-Gao Lai}
\email{Corresponding author: denggaolai@foxmail.com}
\affiliation{Theoretical Quantum Physics Laboratory, RIKEN Cluster for Pioneering Research, Wako-shi, Saitama 351-0198, Japan}

\author{Jie-Qiao Liao}
\email{Corresponding author: jqliao@hunnu.edu.cn}
\affiliation{Key Laboratory of Low-Dimensional Quantum Structures and Quantum Control of Ministry of Education, Key Laboratory for Matter Microstructure and Function of Hunan Province, Department of Physics and Synergetic Innovation Center for Quantum Effects and Applications, Hunan Normal University, Changsha 410081, China}

\begin{abstract}
Quantum squeezing is an important resource in modern quantum technologies, such as quantum precision measurement and continuous-variable quantum information processing. The generation of squeezed states of mechanical modes is a significant task in cavity optomechanics. Motivated by recent interest in multimode optomechanics, it becomes an interesting topic to create quadrature squeezing in multiple mechanical resonators. However, in the multiple-degenerate-mechanical-mode optomechanical systems, the dark-mode effect strongly suppresses the quantum effects in mechanical modes. Here we study the generation of mechanical squeezing in a two-mechanical-mode optomechanical system by breaking the dark-mode effect with the synthetic-gauge-field method. We find that when the mechanical modes work at a finite temperature, the mechanical squeezing is weak or even disappeared due to the dark-mode effect, while the strong mechanical squeezing can be generated once the dark-mode effect is broken. In particular, the thermal-phonon-occupation tolerance of the mechanical squeezing is approximately three orders of magnitude larger than that without breaking the dark-mode effect.  We also generalize  this method to break the dark modes and to create the mechanical squeezing in multiple-mechanical-mode optomechanical systems. Our results describe a general physical mechanism and pave the way towards the generation of noise-resistant quantum resources.
\end{abstract}

\maketitle

\section{Introduction\label{sec:intro}}

Cavity optomechanics~\cite{Kippenberg2008,Meystre2013,Aspelmeyer2014}, which focuses on the radiation-pressure interaction between electromagnetic fields and macroscopic mechanical resonators, has attracted much attention from several subfields in physics, such as optics, quantum physics, and nanosciences. Recently, many notable accomplishments have been made in cavity optomechanics, including the ground-state cooling of mechanical modes~\cite{Wilson-Rae2007,Marquardt2007,Chan2011,Teufel2011a}, the demonstration of strong linearized optomechanical couplings~\cite{Groblacher2009,Sankey2010,Teufel2011b}, and the coherent quantum state transfer between optical and mechanical modes~\cite{Palomaki2013a,Palomaki2013b}. In addition, as a promising research platform, optomechanical systems have been widely used to study the quantum properties of macroscopic mechanical resonators, such as macroscopic quantum superposition~\cite{Bose1997,Marshall2003,Liao2016},   mechanical entanglement~\cite{Riedinger2018,Ockeloen-Korppi2018,Kotler2021,Lepinay2021}, and  mechanical squeezing~\cite{Vinante2013,Pontin2014,Vasilakis2015,Kustura2022}.

In recent decades, the generation of squeezed states in mechanical resonators has become an important goal in cavity optomechanics~\cite{Aspelmeyer2014}, because these states have a wide range of applications in modern quantum technologies, including high-precision measurement~\cite{Caves1980,Abramovici1992,LaHaye2004,Peano2015} and continuous-variable quantum information processing~\cite{Braunstein2005}. Up to now, various schemes have been proposed to generate mechanical squeezing, including parametric amplification~\cite{Agarwal1991,Rugar1991}, quantum measurement~\cite{Clerk2008,Szorkovszky2011,Szorkovszky2013,Meng2020}, parametric modulation~\cite{Mari2009,Liao2011,Pontin2016,Chowdhury2020}, quantum reservoir engineering~\cite{Kronwald2013,Tan2013,Kienzler2015,Huang2021},
mechanical nonlinearities~\cite{Nunnenkamp2010,Asjad2014,Lu2015a,Xiao2017}, and squeezed-state transfer~\cite{Jahne2009,Agarwal2016,Huang2020}. In particular, the mechanical squeezing has been experimentally achieved in electromechanical systems~\cite{Wollman2015,Pirkkalainen2015,Lecocq2015,Lei2016}. Motivated by recent interest in multimode optomechanical systems~\cite{Bhattacharya2008,Xuereb2012a,Xu2013,Nielsen2017,Lai2018,Mercade2021,Lai2021}, an interesting task arising is whether we can simultaneously create single-mode quadrature squeezing of multiple mechanical modes. However, in multiple-mechanical-mode optomechanical systems, the dark-mode effect~\cite{Genes2008,Massel2012,Shkarin2014,Sommer2019,Ockeloen-Korppi2019,Lai2020,Lai2022,Huang2022a,Huang2022b} will be induced when multiple degenerate mechanical modes are coupled in parallel to a common optical mode. It has been shown that the dark-mode effect strongly suppresses both  the ground-state cooling of the mechanical modes and the generation of optomechanical entanglement~\cite{Genes2008,Lai2020,Lai2022,Huang2022a,Huang2022b}. Since the quantum squeezing is also a kind of quantum effect and it is highly susceptible to thermal noise, a natural question now being raised is whether the dark-mode effect will affect the quantum squeezing of the mechanical modes. Furthermore, is it possible to generate and significantly enhance the squeezing of multiple mechanical modes via dark-mode engineering in these multiple-mechanical-mode optomechanical systems?

In this paper, we use the synthetic-gauge-field method to generate strong quantum squeezing of two mechanical modes by breaking the dark-mode effect. Concretely, we consider a two-mechanical-mode optomechanical system involving two degenerate mechanical modes and an optical mode. Here a degenerate optical parametric amplifier (OPA) is placed inside the optical cavity to generate the squeezing of the optical mode~\cite{Wu1986}, then the optical squeezing is transferred to the mechanical modes via the optomechanical interactions working at red-sideband resonance. Note that the degenerate OPA has been widely used in  cavity optomechanics~\cite{Agarwal2016,Qin2018,Qin2022,Huang2009a,Gan2019,Lau2020,Huang2009b,Lu2015b,Xuereb2012b}. For example, the degenerate OPA has been suggested to improve the optomechanical cooling~\cite{Huang2009a,Gan2019,Lau2020}, observe the normal-mode splitting of mechanical resonator~\cite{Huang2009b},  enhance the effective strength of optomechanical interaction~\cite{Lu2015b}, enhance the quantum entanglement~\cite{Xuereb2012b}, and improve the precision of quadrature measurement~\cite{Peano2015}. To generate a synthetic gauge field and break the dark-mode effect, we construct a loop-coupled configuration with the optomechanical and phase-dependent phonon-hopping couplings in this optomechanical system.  The synthetic gauge field has been experimentally achieved in the phase-dependent loop-coupled optomechanical systems~\cite{Schmidt2015,Shen2016,Ruesink2016,Fang2017,Bernier2017,Shen2018,Ruesink2018,Mathew2020,Chen2021}.  We find that quantum squeezing of the two mechanical modes can be largely enhanced by breaking the dark mode, and the generated mechanical squeezing is extraordinarily robust against the thermal noise. In particular, the mechanical squeezing can be persisted even when the thermal phonon number takes an extremely high value. In addition, we generalize the method for generation of mechanical squeezing to the multiple-mechanical-mode optomechanical system by breaking the dark mode with the synthetic-gauge-field method. Our work describes a general physical mechanism, and it provides a different mean to generate the noise-resistant quantum resources against the dark modes.

The rest of this paper is organized as follows. In Sec.~\ref{model}, we describe the physical model and present the Hamiltonian. In Sec.~\ref{Langevineq}, we derive both the quantum
Langevin equations and the Lyapunov equation. In Sec.~\ref{squeezing}, we study the squeezing transfer from the squeezed optical mode to the mechanical modes and investigate the fragile-to-robust mechanical squeezing with the synthetic-gauge-field method. In Sec.~\ref{multiple-mode}, we study the quantum squeezing of multiple mechanical modes in both the dark-mode unbreaking (DMU) and dark-mode breaking (DMB)  cases. In Sec.~\ref{comparison}, we present the comparison on the squeezing generation in previous theoretical proposals and the present proposal. Finally, we give a brief conclusion in Sec.~\ref{conclusion}.

\section{Model and Hamiltonian\label{model}}

%%%%%%%%%%%%%%%%%%%%%%%
\begin{figure}[tbp]
\center
\includegraphics[width=0.35 \textwidth]{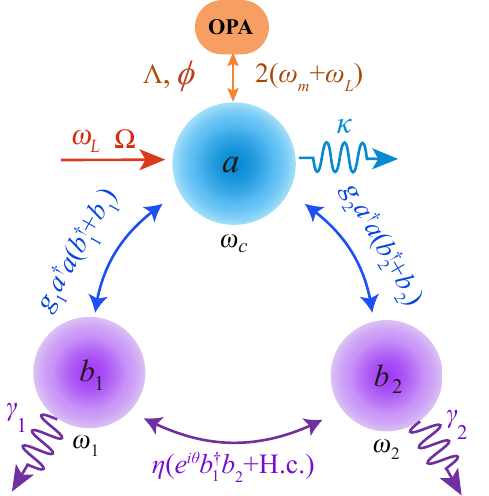}
\caption{Schematic of a loop-coupled three-mode optomechanical system. A degenerate OPA is placed inside the optical cavity and it is pumped by a driving field at frequency $\omega_{m}+\omega_{L}$ with the parametric gain $\Lambda$ and the parametric phase $\phi$. The optical mode $a$ (resonance frequency $\omega_{c}$, decay rate $\kappa$) is optomechanically coupled to two mechanical modes $b_{1}$ ($\omega_{1}$, $\gamma_{1}$) and $b_{2}$ ($\omega_{2}$, $\gamma_{2}$) with the coupling strengths $g_{1}$ and $g_{2}$, respectively. Two mechanical modes are coupled to each other via a phase-dependent phonon-hopping interaction (coupling strength $\eta$,  modulation phase $\theta$). Moreover, the optical cavity is driven by a laser with frequency $\omega_{L}$ and amplitude $\Omega$.}
\label{Fig1}
\end{figure}
%%%%%%%%%%%%%%%%%%%%%%%
We consider a three-mode optomechanical system consisting of an optical mode and two degenerate mechanical modes, as depicted in Fig.~\ref{Fig1}. Here, a degenerate OPA is placed in the optical cavity and used to generate quadrature squeezing of the optical mode.  The generated optical squeezing is transferred to the mechanical mode through the linearized optomechanical couplings working in the red-sideband resonance regime. We assume that the OPA is parametrically pumped by a coherent field at frequency $\omega_{m}+\omega_{L}$ with a gain $\Lambda$ (depending on the pumping intensity) and a phase $\phi$. When the two degenerate mechanical modes are coupled to a common optical mode, the two mechanical modes will be hybridized into a bright mode and a dark mode. In this case, the mechanical dark mode decouples from the optical mode, and it cannot be cooled into its quantum ground state via the optomechanical cooling, then the mechanical squeezing will be destroyed by the residual thermal-excitation noise in the dark mode. To break this mechanical dark mode, we introduce a phase-dependent phonon-hopping interaction between the two mechanical modes with a coupling strength $\eta$ and a modulation phase $\theta$.  Moreover, a driving field with amplitude $\Omega$ and frequency $\omega_{L}$ is applied to the optical cavity.

In a rotating frame defined by $\exp(-i\omega_{L}a^{\dagger}at)$, the system Hamiltonian reads (with $\hbar =1$)
\begin{eqnarray}
H&=&\Delta_{c}a^{\dagger}a+\sum_{l=1,2}[\omega_{l}b_{l}^{\dagger }b_{l}+g_{l}a^{\dagger}a(b_{l}+b_{l}^{\dagger})]+\eta(e^{i\theta}b_{1}^{\dagger}b_{2}+\text{H.c.})\nonumber \\
&&+i\Lambda (e^{i\phi}a^{\dagger 2}e^{-2i\omega_{m}t}-\text{H.c.})+(\Omega a+\text{H.c.}), \label{Hamit1}
\end{eqnarray}
where $a^{\dagger}$ ($a$) and $b^{\dagger}_{l}$ ($b_{l=1,2}$) are, respectively, the creation (annihilation) operators of  the optical and $l$th mechanical modes with the corresponding resonance frequencies $\omega_{c}$ and $\omega_{l}$. The parameter $\Delta_{c}=\omega_{c}-\omega_{L}$  is the detuning of the optical mode frequency with respect to the driving laser frequency. The $g_{l=1,2}$ terms describe the optomechanical couplings between the optical mode and the two mechanical modes, the $\eta$ term describes the phase-dependent phonon-exchange interaction between the two mechanical modes, the $\Omega$ term denotes the driving of the input laser, and the $\Lambda$ term represents  the coupling between the optical mode and the degenerate OPA. Here, the phase-dependent phonon-hopping interaction is introduced to generate a synthetic gauge field, which controls the dark-mode effect.

\section{Quantum Langevin equations and Lyapunov equation \label{Langevineq}}

Based on Hamiltonian~(\ref{Hamit1}), we can obtain the Langevin equations for the optical mode and the two mechanical modes as
\begin{eqnarray}
\label{Langevinab}
\dot{a}&=&-(\kappa+i\Delta_{c})a-i\sum_{l=1,2}[g_{l}a(b_{l}+b_{l}^{\dagger})]+2\Lambda e^{i\phi}a^{\dagger}e^{-2i\omega_{m}t}\nonumber\\
&&-i\Omega+\sqrt{2\kappa}a_{\text{in}},\nonumber\\
\dot{b}_{1}&=&-(\gamma_{1}+i\omega_{1})b_{1}-ig_{1}a^{\dagger}a-i\eta e^{i\theta}b_{2}+\sqrt{2\gamma_{1}}b_{1,\text{in}},\nonumber\\
\dot{b}_{2}&=&-(\gamma_{2}+i\omega_{2})b_{2}-ig_{2}a^{\dagger}a-i\eta e^{-i\theta}b_{1}+\sqrt{2\gamma_{2}}b_{2,\text{in}},
\end{eqnarray}
where $\kappa$ is the decay rate of the optical mode, and $\gamma_{l=1,2}$ is the damping rate of the $l$th mechanical mode. The operators $a_{\text{in}}$ ($a^{\dagger}_{\text{in}}$) and $b_{l,\text{in}}$ ($b^{\dagger}_{l,\text{in}}$) are the noise operators associated with the optical mode and the $l$th mechanical mode, respectively. These noise operators have nonzero correlation functions~\cite{Gardiner2000} $\langle a_{\text{in}}(t) a_{\text{in}}^{\dagger}(t^{\prime})\rangle=\delta(t-t^{\prime})$,  $\langle b_{l,\text{in}}^{\dagger}(t) b_{l,\text{in}}(t^{\prime})\rangle=\bar{n}_{l}\delta(t-t^{\prime})$, and $\langle b_{l,\text{in}}(t) b_{l,\text{in}}^{\dagger}(t^{\prime})\rangle=(\bar{n}_{l}+1)\delta(t-t^{\prime})$,
with $\bar{n}_{l=1,2}$ being the environmental thermal-excitation occupation of the $l$th mechanical mode.

In the strong-driving case, we can expand the operators $o\in\{a,a^{\dagger},b_{l=1,2},b^{\dagger}_{l}\}$ as a sum of their steady-state average values and fluctuation operators (i.e., $o=\langle o\rangle_{ss}+\delta o$). By neglecting the higher-order terms in the fluctuation equations, Eq.~(\ref{Langevinab}) can be linearized as
\begin{eqnarray}
\label{lineLangevin}
\delta\dot{a}&=&-(\kappa +i\Delta )\delta a-i\sum_{l=1,2}[G_{l}(\delta b_{l}+\delta b_{l}^{\dagger})]+\sqrt{2\kappa}a_{\text{in}}\notag\\
&&+2\Lambda e^{i\phi}e^{-2i\omega_{m}t}\delta a^{\dagger}, \notag\\
\delta\dot{b}_{1}&=&-iG_{1}^{\ast}\delta a-(\gamma _{1}+i\omega _{1})\delta b_{1}-i\eta e^{i\theta}\delta b_{2}-iG_{1}\delta a^{\dagger}\notag \\
&&+\sqrt{2\gamma_{1}}b_{1,\text{in}},\notag\\
\delta\dot{b}_{2}&=&-iG_{2}^{\ast }\delta a-i\eta e^{-i\theta }\delta b_{1}-(\gamma _{2}+i\omega _{2})\delta b_{2}-iG_{2}\delta a^{\dagger}\notag \\
&&+\sqrt{2\gamma _{2}}b_{2,\text{in}},
\end{eqnarray}
where $\Delta=\Delta_{c}+2\sum_{l=1,2}g_{l}\text{Re}[\langle b_{l}\rangle_{ss}]$ is the normalized driving detuning with $\text{Re}[\langle b_{l}\rangle_{ss}]$ being the real part of $\langle b_{l}\rangle_{ss}$. The parameter $G_{l=1,2}=g_{l}\langle a\rangle_{ss}$ is the strength of the linearized optomechanical coupling. Note that the steady-state average value $\langle a\rangle_{ss}=-i\Omega/(\kappa+i\Delta)$ is considered to be real by selecting the amplitude $\Omega$.

By introducing the slowly varying fluctuation operators $\delta a=\delta \tilde{a}e^{-i\Delta t}$, $\delta b_{l}=\delta \tilde{b}_{l}e^{-i\omega_{l}t}$,  $a_{\text{in}}=\tilde{a}_{\text{in}}e^{-i\Delta t}$, and $b_{l,\text{in}}=\tilde{b}_{l,\text{in}}e^{-i\omega_{l}t}$, Eq.~(\ref{lineLangevin}) can be written as
\begin{eqnarray}
\label{Langevin}
\delta \dot{\tilde{a}} &=&-\kappa \delta \tilde{a}-i\sum_{l=1,2}G_{l}[\delta\tilde{b}_{l}e^{i(\Delta-\omega_{l})t}+\delta\tilde{b}_{l}^{\dagger }e^{i(\Delta +\omega_{l})t}] \nonumber\\
&&+2\Lambda e^{i\phi}\delta \tilde{a}^{\dagger }e^{2i(\Delta -\omega_{m})t}+\sqrt{2\kappa}\tilde{a}_{\text{in}}, \nonumber\\
\delta \dot{\tilde{b}}_{1} &=&-iG_{1}^{\ast }\delta \tilde{a}e^{-i(\Delta -\omega_{1}) t}-\gamma _{1}\delta \tilde{b}_{1}-i\eta e^{i\theta }\delta \tilde{b}_{2}e^{i(\omega _{1}-\omega_{2})t}\nonumber\\
&&-iG_{1}\delta \tilde{a}^{\dagger}e^{i( \Delta+\omega_{1}) t}+\sqrt{2\gamma_{1}}\tilde{b}_{1,\text{in}}, \nonumber\\
\delta \dot{\tilde{b}}_{2} &=&-iG_{2}^{\ast }\delta \tilde{a}e^{-i(\Delta-\omega_{2}) t}-i\eta e^{-i\theta }\delta \tilde{b}
_{1}e^{-i(\omega _{1}-\omega _{2}) t}-\gamma _{2}\delta \tilde{b}_{2} \nonumber\\
&&-iG_{2}\delta \tilde{a}^{\dagger }e^{i(\Delta +\omega_{2})t}+\sqrt{2\gamma _{2}}\tilde{b}_{2,\text{in}}.
\end{eqnarray}
We point out that these transformed noise operators $\tilde{a}_{\text{in}}$ and $\tilde{b}_{l,\text{in}}$ satisfy the same correlation functions as those of these noise operators $a_{\text{in}}$ and $b_{l,\text{in}}$ before the transformation.

For convenience, we consider the case where the two mechanical modes are degenerate and the driving is in red-sideband resonance  (i.e., $\Delta=\omega_{m}=\omega_{1}=\omega_{2}$). We also assume that the system works in the resolved-sideband limit ($\omega _{l}\gg \kappa$), the quality factors of the mechanical modes are high ($\omega _{l}\gg \gamma _{l}$), and the mechanical frequencies $\omega _{l}$ are much larger than $G_{l}$ and $\Lambda $. Under these conditions, we  make the rotating-wave approximation by discarding the high-frequency oscillating terms. Then Eq.~(\ref{Langevin}) is simplified to
\begin{eqnarray}
\delta \dot{\tilde{a}} &=&-\kappa \delta \tilde{a}-iG_{1}\delta \tilde{b}_{1}-iG_{2}\delta \tilde{b}_{2}+\sqrt{2\kappa }\tilde{a}_{\text{in}
}+2\Lambda e^{i\phi }\delta \tilde{a}^{\dagger }, \nonumber\\
\delta \dot{\tilde{b}}_{1} &=&-iG_{1}\delta \tilde{a}-\gamma_{1}\delta \tilde{b}_{1}-i\eta e^{i\theta }\delta \tilde{b}_{2}+\sqrt{
2\gamma _{1}}\tilde{b}_{1,\text{in}}, \nonumber\\
\delta \dot{\tilde{b}}_{2} &=&-iG_{2}\delta \tilde{a}-i\eta e^{-i\theta }\delta \tilde{b}_{1}-\gamma _{2}\delta \tilde{b}_{2}+\sqrt{
2\gamma _{2}}\tilde{b}_{2,\text{in}}.
\label{Langevinsim}
\end{eqnarray}
We can see from Eq.~(\ref{Langevinsim}) that $\delta \dot{\tilde{a}}$ depends on $\delta \tilde{a}^{\dagger}$,  so the OPA leads to the appearance of the optical squeezing. Moreover, the optical squeezing can be transferred to the two mechanical modes via the linearized optomechanical couplings working at red-sideband resonance.

To investigate quantum squeezing of the two mechanical modes, we introduce the quadrature operators of the optical and mechanical modes
$\delta X_{o}=(\delta \tilde{o}^{\dagger }+\delta \tilde{o})/\sqrt{2}$ and $\delta Y_{o}=i( \delta \tilde{o}^{\dagger }-\delta \tilde{o})/\sqrt{2}$ ($o={a,a^{\dagger},b_{l=1,2},b^{\dagger}_{l}}$), and the corresponding quadrature operators of input quantum noise $X_{o}^{\text{in}}=(\tilde{o}_{\text{in}}^{\dagger}+\tilde{o}_{\text{in}})/\sqrt{2}$ and $Y_{o}^{\text{in}}=i( \tilde{o}_{\text{in}}^{\dagger }-\tilde{o}_{\text{in}})/\sqrt{2}$. Then the linearized Langevin equation~(\ref{Langevinsim}) can be rewritten  as
\begin{eqnarray}
\mathbf{\dot{u}}(t)=\mathbf{Au}(t)+\mathbf{N}(t),\label{MatrixLeq}
\end{eqnarray}
where we introduce the fluctuation operator vector $\mathbf{u}(t)=[\delta X_{b_{1}}, \delta Y_{b_{1}}, \delta X_{b_{2}}, \delta Y_{b_{2}},
\delta X_{a}, \delta Y_{a}]^{T}$ and the noise operator vector $\mathbf{N}(t)=\sqrt{2}[\sqrt{\gamma _{1}}X_{b_{1}}^{\text{in}},\sqrt{\gamma_{1}}Y_{b_{1}}^{\text{in}},\sqrt{\gamma _{2}}X_{b_{2}}^{\text{in}}, \sqrt{\gamma _{2}}Y_{b_{2}}^{\text{in}},\sqrt{\kappa }X_{a}^{\text{in}},\sqrt{\kappa }Y_{a}^{\text{in}}]^{T}$ with the matrix
transpose notation ``$T$". The coefficient matrix is defined by
$\mathbf{A}=\left(\begin{array}{cc}\mathbf{E} & \mathbf{P}\\
-\mathbf{P}^{T} & \mathbf{F} \\\end{array}\right)$,
where we introduce
\begin{eqnarray}
\mathbf{E}=\left(
\begin{array}{cccc}
-\gamma _{1} & 0 & \eta \sin\theta  & \eta \cos\theta \\
0 & -\gamma _{1} & -\eta \cos \theta  & \eta \sin \theta \\
-\eta \sin \theta  & \eta \cos \theta  & -\gamma _{2} & 0 \\
-\eta \cos \theta  & -\eta \sin \theta  & 0 & -\gamma _{2}
\end{array}
\right)
\label{Ematrix},
\end{eqnarray}
and
\begin{eqnarray}
\mathbf{F}=\left(
\begin{array}{cc}
-(\kappa-2\Lambda\cos\phi)  & 2\Lambda\sin\phi \\
2\Lambda\sin\phi  &  -(\kappa+2\Lambda\cos\phi)
\end{array}
\right).
\label{Fmatrix}
\end{eqnarray}
In addition, the matrix $\mathbf{P}$ is defined by these nonzero elements:
$\mathbf{P}_{12}=G_{1}$, $\mathbf{F}_{21}=-G_{1}$, $\mathbf{P}_{32}=G_{2}$, and $\mathbf{P}_{41}=-G_{2}$.
The stability condition of this system can be obtained by confirming that all the eigenvalues of the coefficient matrix $\mathbf{A}$ have negative real parts. In our calculations, we used proper parameters to confirm the stability of the system with the Routh-Hurwitz criterion~\cite{Gradshteyn2014}.

To study the steady-state squeezing of the two mechanical modes, we calculate the steady-state value of the covariance matrix $\mathbf{V}$, defined by the matrix elements
\begin{equation}
\mathbf{V}_{ij}=\frac{1}{2}[\langle \mathbf{u}_{i}(\infty) \mathbf{u}_{j}(\infty ) \rangle +\langle \mathbf{u}_{j}( \infty)\mathbf{u}_{i}(\infty )\rangle], \hspace{0.3 cm}i,j=1\text{-}6. \label{covarimatrix}
\end{equation}
In the linearized optomechanical system, the covariance matrix $\mathbf{V}$ obeys the Lyapunov equation~\cite{Vitali2007}
\begin{equation}
\mathbf{A}\mathbf{V}+\mathbf{V}\mathbf{A}^{T}=-\mathbf{Q}, \label{Lyapunov}
\end{equation}
where the matrix $\mathbf{Q}$ is defined by $\mathbf{Q}=(\mathbf{C}+\mathbf{C}^{T})/2$, with $\mathbf{C}$ being the noise correlation matrix defined by the matrix elements
$\langle \mathbf{N}_{k}(s)\mathbf{N}_{l}(s^{\prime})\rangle=\mathbf{C}_{k,l}\delta (s-s^{\prime })$. Thus, we obtain the diagonal matrix $\mathbf{Q}=\mathrm{diag} \{\gamma_{1}(2\bar{n}_{1}+1),\gamma_{1}(2\bar{n}_{1}+1),\gamma_{2}(2\bar{n}_{2}+1),\gamma_{2}(2\bar{n}_{2}+1),\kappa,\kappa\}$.

\section{Quadrature squeezing \label{squeezing}}
In this section, we present the definition of the degree of squeezing,  study the  squeezing transfer from the optical mode to the two mechanical modes, and  investigate the fragile-to-robust mechanical squeezing with the synthetic-gauge-field method. Concretely, we first calculate the degree of squeezing of both the optical mode and mechanical modes when the mechanical modes are connected with zero-temperature environments. Then, we study the mechanical squeezing  in both the absence and presence of the synthetic magnetism when the mechanical modes are connected with nonzero-temperature environments.

\subsection{Squeezing  transfer from squeezed optical mode to mechanical mode  \label{Gensqueezing}}

In this scheme, due to the zero expectations of $\langle\delta X_{o}\rangle$ and $\langle\delta Y_{o}\rangle$ ($o= a$ and $b_{l=1,2}$), the quadrature squeezing can be measured by either the mean square fluctuations $\langle\delta X^{2}_{o}\rangle$ or $\langle\delta Y^{2}_{o}\rangle$, which is just one of the six diagonal elements ($\mathbf{V}_{jj}$ for $j=1\text{-}6$ ) of the covariance matrix $\mathbf{V}$ defined in Eq.~(\ref{covarimatrix}). According to the Heisenberg uncertainty principle, the product of $\langle\delta X^{2}_{o}\rangle$ and $\langle\delta Y^{2}_{o}\rangle$ obeys the inequality
\begin{equation}
\langle\delta X^{2}_{o}\rangle\langle\delta Y^{2}_{o}\rangle\geq\frac{1}{4}\vert[\delta X_{o},\delta Y_{o}]\vert^{2}=\frac{1}{4},  \label{inequality}
\end{equation}
where we used the relation $[\delta X_{o},\delta Y_{o}]=i$. Therefore, if either $\langle\delta X^{2}_{o}\rangle$ or $\langle\delta Y^{2}_{o}\rangle$ is below $1/2$, the corresponding mode will exhibit quadrature squeezing. The degree of squeezing is defined by~\cite{Agarwal2016}
\begin{equation}
\mathcal{S}_{Z}=-10\log_{10}\frac{\langle Z^{2}\rangle}{\langle Z^{2}\rangle_{\text{zpf}}} \label{squeezdegree}
\end{equation}
with $Z=\delta X_{o}$ or $\delta Y_{o}$, where $\langle Z^{2}\rangle_{\text{zpf}}$ denotes the zero-point fluctuation of the operator $Z$. The degree of squeezing $\mathcal{S}_{Z}>0$ dB implies that the corresponding mode is squeezed.

%%%%%%%%%%%%%%%%%%%%%%%
\begin{figure}[tbp]
\center
\includegraphics[width=0.48 \textwidth]{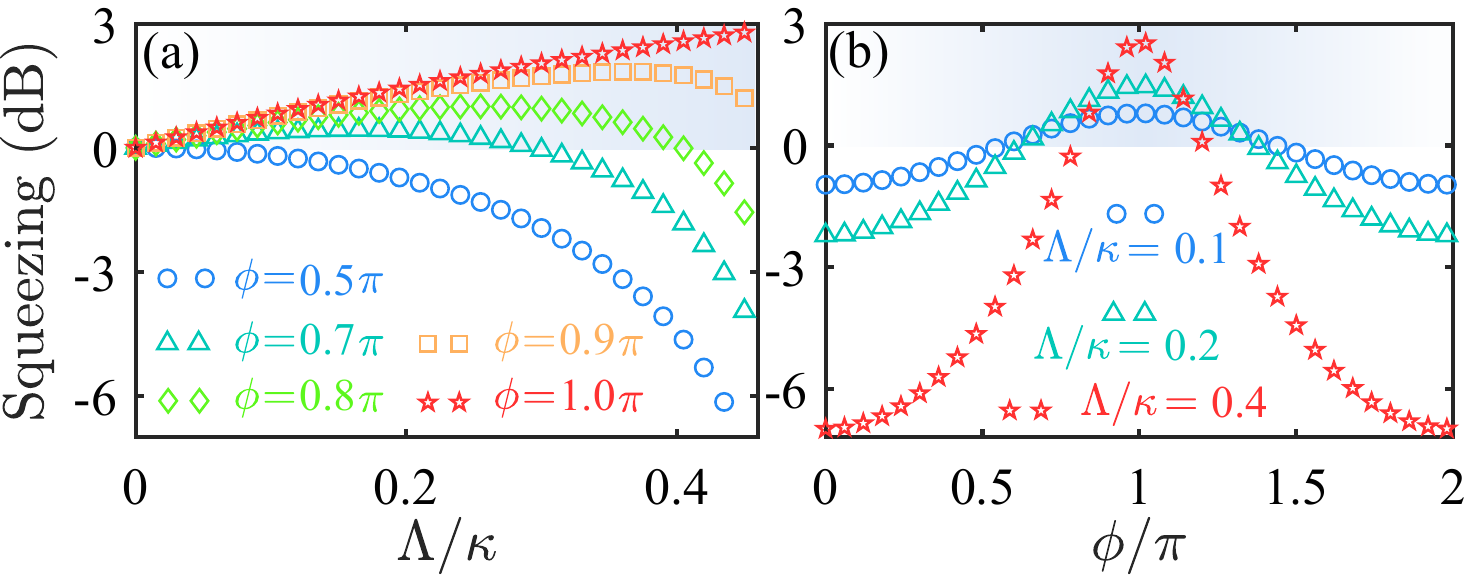}
\caption{Degree of the squeezing $\mathcal{S}_{\delta X_{a}}$ versus (a) the scaled gain $\Lambda/\kappa$ when the  phase $\phi/\pi$ takes various values and versus (b) the phase $\phi/\pi$ when the gain $\Lambda/\kappa$ takes various values. Other parameters are $\omega_{1}/\kappa=\omega_{2}/\kappa=10$, $\gamma_{1}/\kappa=\gamma_{2}/\kappa=10^{-5}$, $G_{1}/\kappa=G_{2}/\kappa=0$, $\bar{n}_{1}=\bar{n}_{2}=0$, $\eta/\kappa=0$, and $\theta=0$.}
\label{Fig2}
\end{figure}
%%%%%%%%%%%%%%%%%%%%%%%

The degenerate OPA is a good candidate for the generation of the optical squeezing, so it is interesting to study the influence of both the parametric gain $\Lambda$ and phase $\phi$ on the optical squeezing. In Figs.~\ref{Fig2}(a) and~\ref{Fig2}(b), we show the degree of squeezing $\mathcal{S}_{\delta X_{a}}$ as a function of the scaled gain $\Lambda/\kappa$ and phase $\phi/\pi$, respectively. We note that the optical mode is not squeezed ($\mathcal{S}_{\delta X_{a}}=0$ dB) in the absence of the OPA ($\Lambda=0$). However, $\mathcal{S}_{\delta X_{a}}>0$ dB appears when the OPA is introduced ($\Lambda>0$), which means that the optical squeezing appears. Especially, the optimal optical squeezing occurs at $\phi=\pi$, and it becomes stronger as the parametric gain $\Lambda$ increases. These results indicate that the optical squeezing can be generated when the degenerate OPA is placed inside the optical cavity.

\begin{figure}[tbp]
\center
\includegraphics[width=0.48 \textwidth]{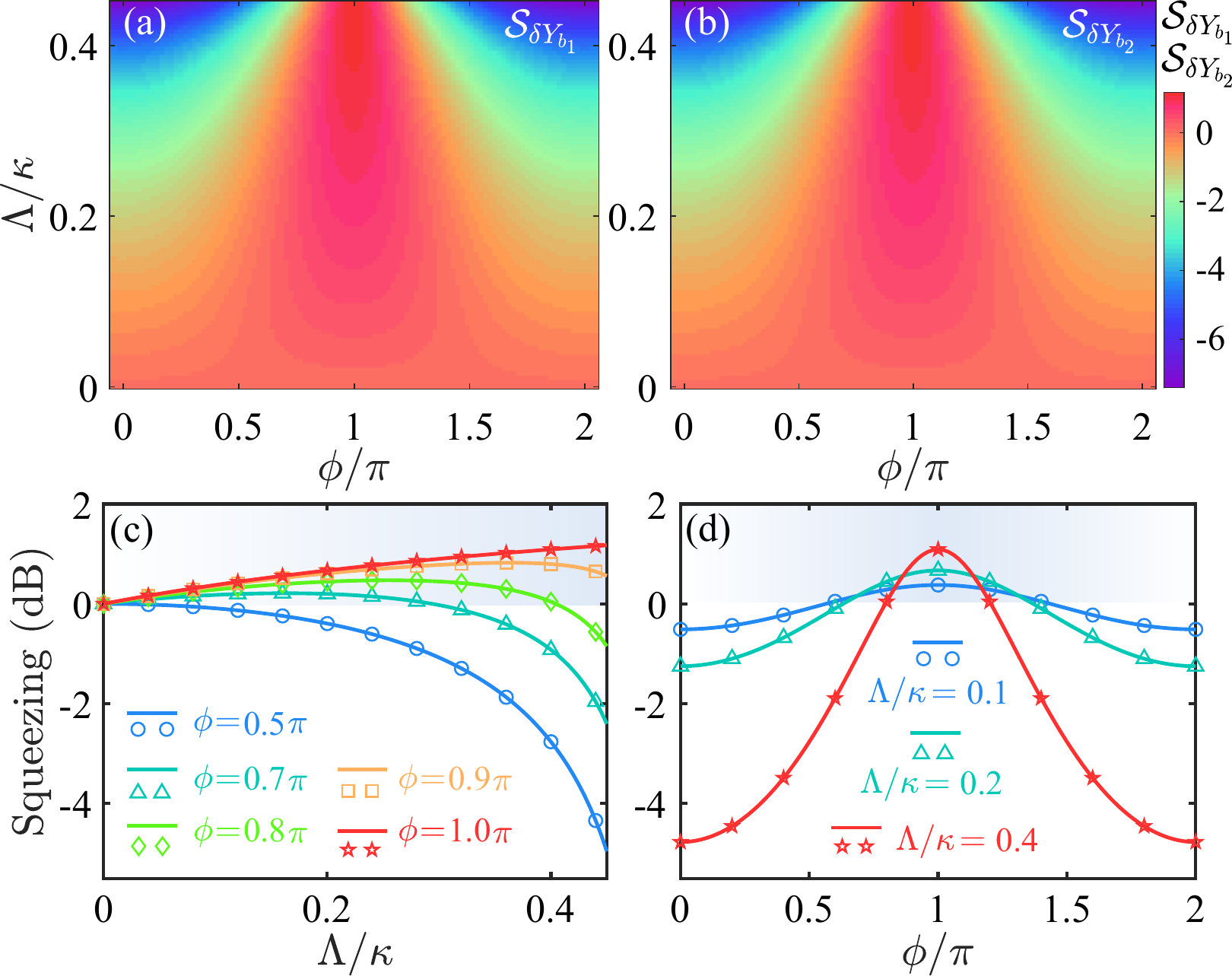}
\caption{Degrees of the squeezing (a) $\mathcal{S}_{\delta Y_{b_{1}}}$ and (b) $\mathcal{S}_{\delta Y_{b_{2}}}$ versus the scaled  gain $\Lambda/\kappa$ and the phase $\phi/\pi$. (c) $\mathcal{S}_{\delta Y_{b_{1}}}$ and $\mathcal{S}_{\delta Y_{b_{2}}}$ versus  $\Lambda/\kappa$ when  $\phi/\pi$ takes various values. (d) $\mathcal{S}_{\delta Y_{b_{1}}}$ and $\mathcal{S}_{\delta Y_{b_{2}}}$ versus $\phi/\pi$ when $\Lambda/\kappa$ takes various values. Note that the solid curves and the symbols represent the degrees of the squeezing $\mathcal{S}_{\delta Y_{b_{1}}}$ and $\mathcal{S}_{\delta Y_{b_{2}}}$, respectively. Other parameters are $\omega_{1}/\kappa=\omega_{2}/\kappa=10$, $\gamma_{1}/\kappa=\gamma_{2}/\kappa=10^{-5}$, $G_{1}/\kappa=G_{2}/\kappa=0.1$, $\bar{n}_{1}=\bar{n}_{2}=0$, $\eta/\kappa=0$, and $\theta=0$.}
\label{Fig3}
\end{figure}

Since the degenerate OPA plays an important role in the appearance of the mechanical squeezing, it is necessary to investigate the dependence of the mechanical squeezing on the parametric gain $\Lambda$ and phase $\phi$. In Figs.~\ref{Fig3}(a) and ~\ref{Fig3}(b), we show the degrees of squeezing $\mathcal{S}_{\delta Y_{b_{1}}}$ and $\mathcal{S}_{\delta Y_{b_{2}}}$ as functions of the scaled gain $\Lambda/\kappa$ and phase $\phi/\pi$. We can see that the degree of squeezing $\mathcal{S}_{\delta Y_{b_{1}}}$ ($\mathcal{S}_{\delta Y_{b_{2}}}$) reaches the maximum value at the optimal phase $\phi=\pi$ when the parametric gain $\Lambda/\kappa$ takes the maximum value. We also see that $\mathcal{S}_{\delta Y_{b_{1}}}$ and $\mathcal{S}_{\delta Y_{b_{2}}}$ can be significantly enhanced with the increase of the gain $\Lambda/\kappa$ around the optimal phase $\phi=\pi$. These features can be seen more clearly in Figs.~\ref{Fig3}(c) and ~\ref{Fig3}(d).  We observe that the two mechanical modes are not squeezed ($\mathcal{S}_{\delta Y_{b_{1}}}=\mathcal{S}_{\delta Y_{b_{2}}}=0$ dB) in the absence of the OPA ($\Lambda=0$). However, once the OPA is introduced ($\Lambda>0$), the mechanical squeezing appears ($\mathcal{S}_{\delta Y_{b_{l}}}>0$ dB) when we take appropriate parametric phases $\phi$ (see the upper part of the panel). In particular, the change of the mechanical squeezing with parameters is consistent with those of optical squeezing, and the maximal degree of mechanical squeezing is significantly less than that of optical squeezing.  These results indicate that the squeezing of the optical mode has been transferred to the mechanical mode,  while there exists some loss in the transfer process~\cite{Agarwal2016,Huang2020}.

The underlying physics for the generation of the mechanical squeezing can be explained as follows. In the absence of the OPA, neither optical nor mechanical modes can be squeezed under the current parameter conditions. However, when the OPA is placed in the optical cavity, the squeezing of the optical mode can be created, and the optical squeezing can be further transferred into the mechanical mode via the linearized optomechanical interaction working in the red-side resonance. Thus, the OPA provides the physical origin for generating the quantum squeezing of the two mechanical modes.

%%%%%%%%%%%%%%%%%%%%%%%
%%%%%%%%%%%%%%%%%%%%%%%

\begin{figure}[tbp]
\center
\includegraphics[width=0.47 \textwidth]{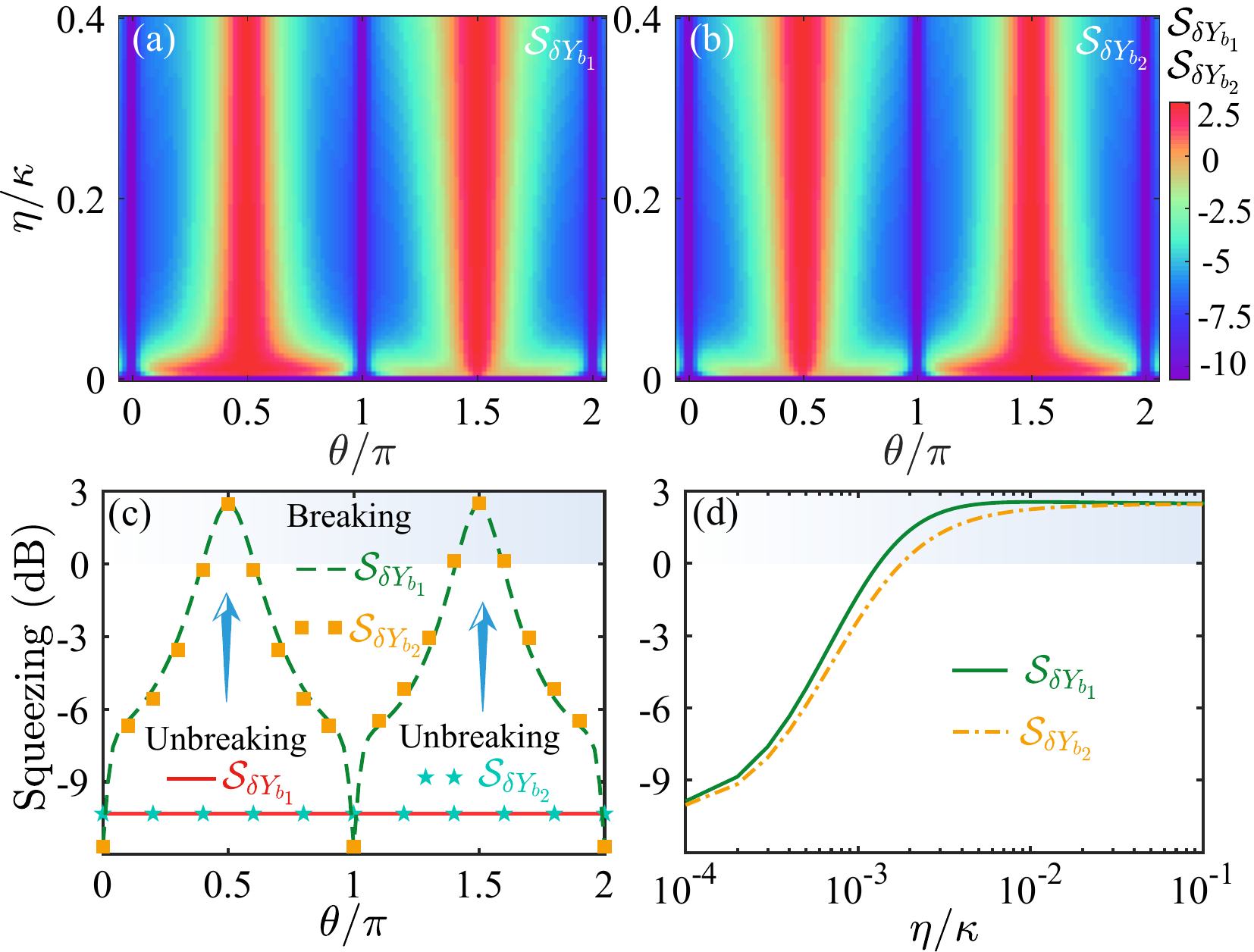}
\caption{Degrees of the squeezing (a) $\mathcal{S}_{\delta Y_{b_{1}}}$ and (b) $\mathcal{S}_{\delta Y_{b_{2}}}$ versus the scaled phonon-hopping coupling strength $\eta/\kappa$ and the modulation phase $\theta/\pi$. (c) $\mathcal{S}_{\delta Y_{b_{1}}}$ and $\mathcal{S}_{\delta Y_{b_{2}}}$ versus  $\theta/\pi$ in both the DMU and DMB cases. (d) $\mathcal{S}_{\delta Y_{b_{1}}}$ and $\mathcal{S}_{\delta Y_{b_{2}}}$ versus the scaled phonon-hopping coupling strength $\eta/\kappa$ when $\theta=\pi/2$. Other parameters are $\omega_{1}/\kappa=\omega_{2}/\kappa=10$, $\gamma_{1}/\kappa=\gamma_{2}/\kappa=10^{-5}$, $G_{1}/\kappa=G_{2}/\kappa=0.1$, $\bar{n}_{1}=\bar{n}_{2}=10$, $\Lambda/\kappa=0.45$, and $\phi=\pi$.}
\label{Fig4}
\end{figure}

In the two-mechanical-mode optomechanical system, when the two mechanical modes are degenerate, there exists a mechanical dark mode. It has been shown that the dark mode will suppress the ground-state cooling of the mechanical modes and the optomechanical entanglement~\cite{Genes2008,Lai2020,Lai2022,Huang2022a,Huang2022b}. Since the quantum squeezing is a kind of quantum effect, it is interesting to study the quantum squeezing behaviors in the system in both the absence ($\eta=0$) and presence ($\eta\neq0$) of the synthetic magnetism. As shown in Fig.~\ref{Fig4}, we plot the degrees of squeezing $\mathcal{S}_{\delta Y_{b_{1}}}$ and $\mathcal{S}_{\delta Y_{b_{2}}}$ as functions of the scaled phonon-hopping coupling strength $\eta/\kappa$ and the modulation phase $\theta/\pi$. From Figs.~\ref{Fig4}(a), \ref{Fig4}(b), and \ref{Fig4}(c) we can see that the mean square fluctuations $\langle\delta Y^{2}_{b_{1}}\rangle$ and $\langle\delta Y^{2}_{b_{2}}\rangle$ are squeezed in a large range of $\eta$ when the phase is around $\theta=\pi/2$ and $\theta=3\pi/2$.  Meanwhile, the largest quantum squeezing of each mechanical mode emerges at $\theta=\pi/2$ and $\theta=3\pi/2$, which are related to the strongest quantum interference between the two excitation-transport channels. Moreover, the mechanical squeezing is completely lost at $\theta=n\pi$ (for an integer $n$), which corresponds to the emergence of the dark-mode effect.  In addition, the degree of squeezing $\mathcal{S}_{\delta Y_{b_{1}}}$ ($\mathcal{S}_{\delta Y_{b_{2}}}$) is slightly larger than $\mathcal{S}_{\delta Y_{b_{2}}}$ ($\mathcal{S}_{\delta Y_{b_{1}}}$) in the region $0<\theta<\pi$ ($\pi<\theta<2\pi$). This phenomenon shows that the periodically controllable and switchable quantum squeezing of each mechanical mode can be performed by just tuning the modulation phase $\theta$. In Fig.~\ref{Fig4}(d), we see that the squeezing of two mechanical modes can be generated and significantly enhanced by properly increasing the phonon-hopping coupling strength $\eta/\kappa$ when $\theta=\pi/2$. The results indicate that the synthetic magnetism plays an important role in the enhancement of mechanical squeezing.

%%%%%%%%%%%%%%%%%%%%%%%
%%%%%%%%%%%%%%%%%%%%%%%

We now explain the physical mechanism of the squeezing enhancement based on the DMB effect. In the presence of the synthetic magnetism ($\eta\neq0$), we introduce two new normal modes associated with the synthetic magnetism as $\tilde{B}_{\pm}= f\delta \tilde{b}_{1(2)}\mp e^{\pm i\theta}h\delta \tilde{b}_{2(1)}$. Based on Eq.~(\ref{Langevinsim}),  the linearized Hamiltonian can be rewritten as
\begin{eqnarray}
H_{\text{lin}}&=&\sum_{j=\pm}(\tilde{G}_{j}\delta \tilde{a}^{\dagger}\tilde{B}_{j}+\text{H.c.})+i\Lambda (e^{i\phi}\delta \tilde{a}^{\dagger 2}-e^{-i\phi}\delta \tilde{a}^{2}),\label{DigHamitlin}
\end{eqnarray}
where we introduce the two coupling strengths $\tilde{G}_{\pm}=fG_{1(2)}\mp e^{\mp i\theta}hG_{2(1)}$, with  $f=|\omega _{f}|/(\omega _{f}^{2}+\eta^{2})^{1/2}$, $h=\eta f/\omega_{f}$, and $\omega_{f}=\{\omega _{2}-\omega _{1}-[(\omega_{1}-\omega _{2})^{2}+4\eta ^{2}]^{1/2}\}/2$. In the degenerate-resonator ($\omega_{1}=\omega_{2}$) and symmetric-coupling ($G_{1}=G_{2}=G$) case, the coupling strengths $\tilde{G}_{\pm}$ can be simplified as
\begin{eqnarray}
\label{DDigGequwG}
\tilde{G}_{+}=G(1+e^{-i\theta})/\sqrt{2},\hspace{0.5 cm}\tilde{G}_{-}=G(1-e^{i\theta})/\sqrt{2}.
\end{eqnarray}
 Equation~(\ref{DDigGequwG}) shows that one of the two mechanical normal modes decouples from the optical mode and becomes a dark mode when $\theta=n\pi$ for an integer $n$. In this case, the mechanical modes cannot be cooled into their quantum ground states via the optomechanical-cooling channel. In general cases of $\theta\neq n\pi$, a coupling between the dark mode and the optical mode can be realized, i.e., both the two normal modes $\tilde{B}_{\pm}$ are coupled to the optical mode $a$, which indicates that the dark mode is broken and then the ground-state cooling of the two mechanical modes can be realized. Physically, a reconfigurable synthetic gauge field can be formed by modulating $\theta$, and this enables a flexible switching between the DMB and DMU cases.

Combined with the above DMB physical mechanism, the physical reason for the generation of mechanical squeezing shown in Fig.~\ref{Fig4} can be explained as follows. In the absence of the synthetic magnetism ($\eta=0$), this system can be described by the two degenerate mechanical modes coupled to a common optical mode, then the two mechanical modes will be hybridized into a bright mode and a dark mode. The dark mode is decoupled from the optical mode, then the thermal excitations concealed in the dark mode cannot be extracted by the cooling channel associated with the optical mode. Hence, the dark mode cannot be cooled to its quantum ground state~\cite{Cooling}. Furthermore, the residual thermal noise in the dark mode will fully destroy the mechanical squeezing transferred from the squeezed optical mode. However, in the presence of the synthetic magnetism ($\eta\neq0$), a coupling between the optical mode and the dark mode can be realized by modulating the phase $\theta$, which indicates that this dark mode is broken. In this case, the two mechanical modes can be cooled into their quantum ground states~\cite{Cooling}, and hence the mechanical squeezing transferred from the squeezed optical mode can be created. Thus, the synthetic magnetism not only enables a flexible switching between the DMB and DMU regimes, but also provides a clear perspective for generating dark-mode-immune quantum resources in multiple-mechanical-mode optomechanical systems.

\subsection{Fragile-to-robust squeezing \label{Ftrsqueezing}}

\begin{figure}[tbp]
\center
\includegraphics[width=0.47 \textwidth]{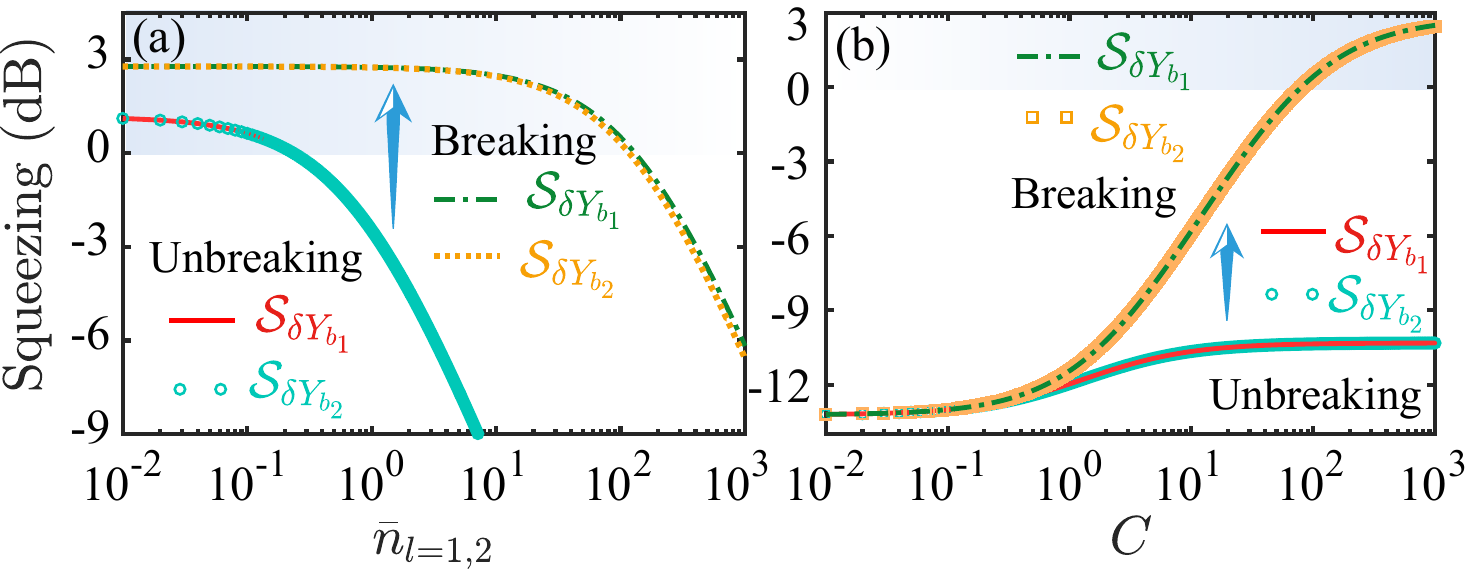}
\caption{Degrees of the squeezing $\mathcal{S}_{\delta Y_{b_{1}}}$ and $\mathcal{S}_{\delta Y_{b_{2}}}$ versus (a) the thermal phonon numbers $\bar{n}_{l=1,2}$ and (b) the optomechanical cooperativity parameter $C$ in both the DMU and DMB cases. Note that the optomechanical coupling strengths $G_{1}/\kappa=G_{2}/\kappa=0.1$ are used in panel (a)  and the thermal phonon numbers $\bar{n}_{l=1,2}=10$ are used in panel (b). Other parameters are $\omega_{1}/\kappa=\omega_{2}/\kappa=10$, $\gamma_{1}/\kappa=\gamma_{2}/\kappa=10^{-5}$, $\Lambda/\kappa=0.45$, $\phi=\pi$, $\eta/\kappa=0.1$, and $\theta=\pi/2$.}
\label{Fig5}
\end{figure}

In the presence of the dark-mode effect ($\eta=0$), the quantum squeezing of two mechanical modes is highly susceptible to thermal noise. To study whether the synthetic-gauge-field method can generate the robust squeezing of mechanical modes, we plot the degrees of squeezing $\mathcal{S}_{\delta Y_{b_{1}}}$ and $\mathcal{S}_{\delta Y_{b_{2}}}$ as functions of the mechanical thermal phonon numbers $\bar{n}_{l=1,2}$ in both the DMB and DMU cases. As shown in Fig.~\ref{Fig5}(a), the squeezing of mean square fluctuations ($\mathcal{S}_{\delta Y_{b_{1}}}>0$ dB and $\mathcal{S}_{\delta Y_{b_{2}}}>0$ dB) emerges only when the thermal phonon numbers $\bar{n}_{l=1,2}\leq0.1$ (see the point $10^{-1}$ on the horizontal axis) in the DMU case. This indicates that the quantum squeezing of the two mechanical modes does not exist in this system when the mechanical modes are not cooled into their quantum ground states. In the DMB case, differently, the degrees of squeezing $\mathcal{S}_{\delta Y_{b_{1}}}>0$ dB and $\mathcal{S}_{\delta Y_{b_{2}}}>0$ dB can persist even when the thermal phonon numbers are about $\bar{n}_{l=1,2}\approx100$ (see the point $10^{2}$ on the horizontal axis), and it is about three orders of magnitude larger than the case where the dark-mode effect is not broken. For the typical mechanical resonators with the resonance frequencies of $\omega_{1}=\omega_{2}\approx2\pi\times10^{8}$ Hz,  the corresponding temperature of the environment can be estimated as $T_{l=1,2}\approx482$ mK when $\bar{n}_{l=1,2}\approx100$.

The physical reason behind this phenomenon can be understood in this way. In the absence of the synthetic magnetism ($\eta=0$) and when the two mechanical modes work at nonzero temperatures ($\bar{n}_{l=1,2}\neq0$), the thermal excitations concealed in the mechanical dark mode cannot be extracted via the optical mode, thus the mechanical squeezing transferred from the squeezed optical mode is fully destroyed by the thermal noise. However, when the synthetic magnetism is introduced, the dark-mode effect can be broken by modulating the phase. As a result, the two mechanical modes can be cooled near to their quantum ground states, and hence the mechanical squeezing can be generated. This DMB physical mechanism switches the tolerance of the mechanical squeezing to the thermal noise from extremely fragile to extraordinary robust.

%%%%%%%%%%%%%%%%%%%%%%%
%%%%%%%%%%%%%%%%%%%%%%%

The optomechanical cooperativity parameter $C=|G^{2}|/\kappa\gamma$ (depending on the pump power of optical cavity) is an important physical quantity for the generation of mechanical squeezing. To investigate the influence of the parameter $C$ on the generation of mechanical squeezing  when the mechanical modes work at nonzero-temperature environment, we plot the degrees of squeezing $\mathcal{S}_{\delta Y_{b_{1}}}$ and $\mathcal{S}_{\delta Y_{b_{2}}}$ as functions of the parameter $C$ in both the DMU and DMB cases. Figure~\ref{Fig5}(b) shows that the mechanical squeezing does not appear ($\mathcal{S}_{\delta Y_{b_{1}}}<0$ dB and $\mathcal{S}_{\delta Y_{b_{2}}}<0$ dB) in the DMU case. However,  when the dark mode is broken, the degrees of squeezing $\mathcal{S}_{\delta Y_{b_{1}}}$ and $\mathcal{S}_{\delta Y_{b_{2}}}$ increase with the increase of $C$, thus we can obtain the strong mechanical squeezing by increasing the parameter $C$ ($C=1000$ corresponds to $G/\kappa=0.1$). The results indicate that the strength of mechanical squeezing is related to the optomechanical coupling strength.  However, when the mechanical modes work at the nonzero-temperature environment, DMB is the dominant factor for the  generation of squeezing, and the strong mechanical squeezing can be obtained only when the dark mode is broken.

\section{Mechanical squeezing in a multiple-mechanical-mode optomechanical system \label{multiple-mode}}

In Sec.~\ref{Gensqueezing}, we have studied the generation of quantum squeezing of two mechanical modes by breaking the dark-mode effect with the synthetic-gauge-field method. We now generalize this squeezing-generation method to a multiple-mechanical-mode optomechanical system consisting of an optical mode and $N$ ($N\geq3$) mechanical modes. Here, the optical mode and $N$ mechanical modes are coupled via the optomechanical interactions. The nearest-neighbor mechanical modes are coupled via the phase-dependent phonon-exchange interactions. In a rotating frame defined by $\exp(-i\omega_{L}a^{\dagger}at)$, the system Hamiltonian reads ($\hbar =1$)
\begin{eqnarray}
\label{HamitN}
H&=&\Delta_{c}a^{\dagger}a+\sum_{l=1}^{N}[\omega_{l}b_{l}^{\dagger }b_{l}+g_{l}a^{\dagger}a(b_{l}+b_{l}^{\dagger})]+(\Omega a+\text{H.c.})\nonumber \\
&&+\sum_{l=1}^{N-1}[\eta_{l}(e^{i\theta_{l}}b_{l}^{\dagger}b_{l+1}+\text{H.c.})]+i\Lambda (e^{i\phi}a^{\dagger 2}e^{-2i\omega_{m}t}-\text{H.c.}), \nonumber \\
\end{eqnarray}
where $b_{l}^{\dagger }$ and $b_{l}$ are, respectively, the creation and annihilation operators of the $l$th mechanical mode with resonance frequency $\omega_{l}$. The $g_{l}$ terms denote the optomechanical couplings between the optical mode and the $l$th mechanical mode. The $\eta_{l}$ terms describe the phase-dependent phonon-exchange interactions between the two nearest-neighbor mechanical modes with the modulation phase $\theta_{l}$. Other operators and parameters have been defined in Eq.~(\ref{Hamit1}).

By using the same method as  shown  in Sec.~\ref{Langevineq}, we can obtain the linearized Hamiltonian of this multiple-mechanical-mode optomechanical system. We know that in the absence of synthetic magnetism ($\eta_{l}=0$), the $N$ ($N\geq3$) mechanical modes will be hybridized into a bright mode $B=\sum_{l=1}^{N}\delta \tilde{b}_{l}/\sqrt{N}$ and $(N-1)$ dark modes~\cite{Lai2020,Lai2022}.  When the synthetic magnetism is introduced into this system (i.e., $\eta_{l}\neq0$), all the dark modes can be broken by adjusting the phase $\theta_{1}\neq2n\pi$~\cite{Lai2020,Lai2022}. Thus, this DMB physical mechanism provides a possibility to switch the DMU case to the DMB case  in this multiple-mechanical-mode optomechanical system.

%%%%%%%%%%%%%%%%%%%%%%%
\begin{figure}[tbp]
\center
\includegraphics[width=0.47 \textwidth]{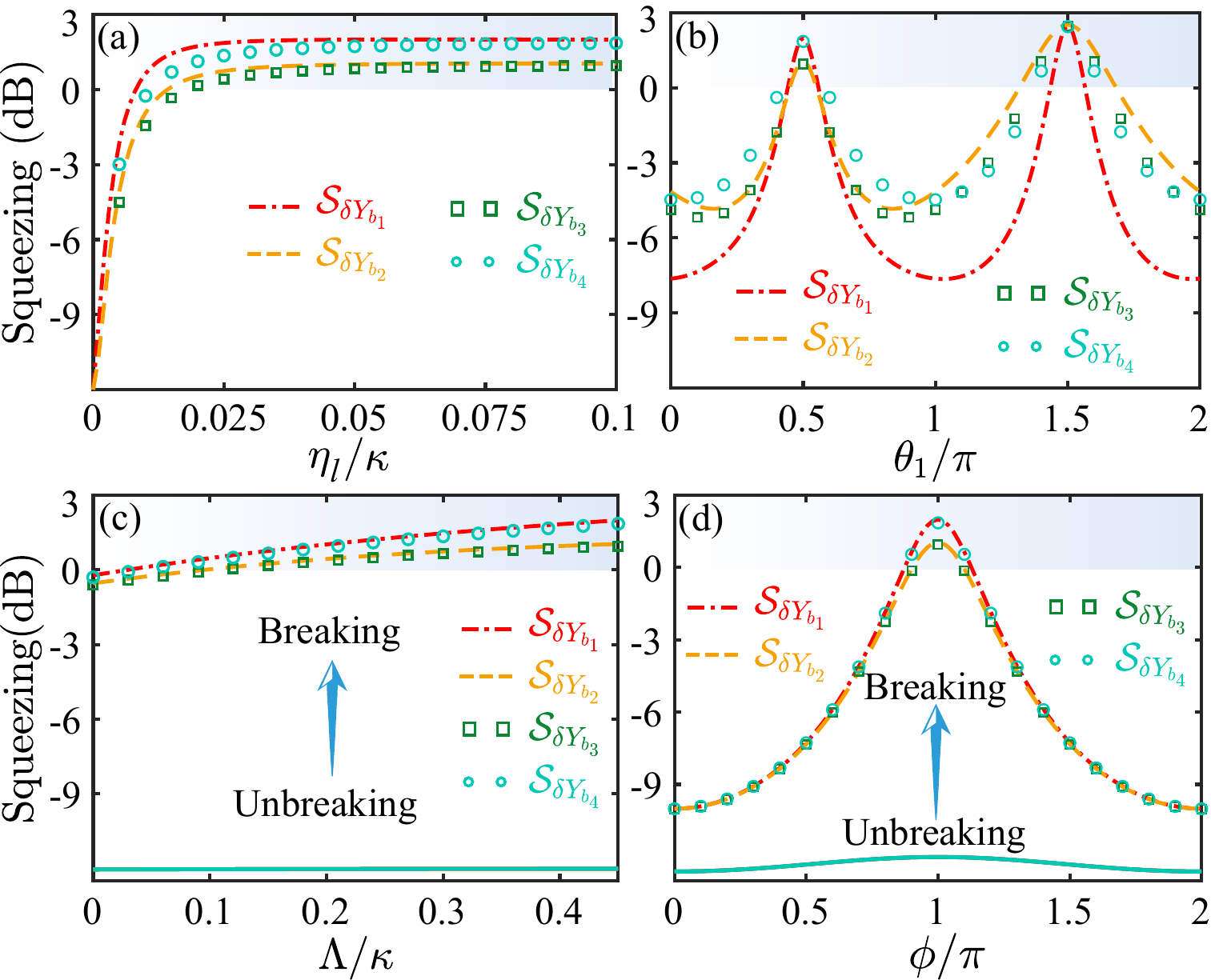}
\caption{Degrees of the squeezing  $\mathcal{S}_{Y_{b_{l}}}$ for $l=1\text{-}4$ versus (a) the scaled phonon-hopping coupling strength $\eta_{l}/\kappa$ and (b) the modulation phase $\theta_{1}/\pi$ when $\Lambda/\kappa=0.45$  and $\phi=\pi$.  Note that parameter $\theta_{l}=\pi/2$ is used in (a) and parameters $\theta_{l=2,3,4}=\pi/2$ and $\eta_{l}/\kappa=0.1$ are used in (b). $\mathcal{S}_{Y_{b_{l}}}$ versus (c) the gain $\Lambda/\kappa$ and (d) the phase $\phi/\pi$ when $\eta_{l}/\kappa=0.1$ and $\theta_{l}=\pi/2$ in both the DMU and DMB cases.  Parameter $\phi=\pi$ is used in (c) and parameter $\Lambda/\kappa=0.45$ is used in (d). Other parameters are $\omega_{l}/\kappa=10$, $\gamma_{l}/\kappa=10^{-5}$, $G_{l}/\kappa=0.1$, and $\bar{n}_{l}=10$.}
\label{Fig6}
\end{figure}
%%%%%%%%%%%%%%%%%%%%%%%

To clearly investigate the influence of  synthetic magnetism on the quantum squeezing of these $N$ mechanical modes, in Figs.~\ref{Fig6}(a) and ~\ref{Fig6}(b) we plot the degrees of squeezing $\mathcal{S}_{Y_{b_{l}}}$ for $l=1\text{-}4$ as functions of the phonon-hopping coupling strengths $\eta_{l}/\kappa$ and the modulation phase $\theta_{1}/\pi$. Here we can see that the squeezing of four mechanical modes can be generated and significantly enhanced by properly increasing the phonon-hopping coupling strength $\eta_{l}/\kappa$, which means that all the dark modes are broken. Similar to the case of squeezing of two mechanical modes, the largest squeezing of each mechanical mode emerges at $\theta_{1}=\pi/2$ and $\theta_{1}=3\pi/2$, which are related to the strongest quantum interference effect.

To study the effect of the degenerate OPA on the quantum squeezing of $N$ mechanical modes, in Figs.~\ref{Fig6}(c) and ~\ref{Fig6}(d) we plot the degrees of squeezing $\mathcal{S}_{Y_{b_{l}}}$ for $l=1\text{-}4$ as functions of the scaled gain $\Lambda/\kappa$ and phase $\phi/\pi$ . Here we can see that the  squeezing of four mechanical modes does not appear ($\mathcal{S}_{Y_{b_{l}}}\leq0$ dB) in the absence of the OPA ($\Lambda=0$),  and the mechanical squeezing can be generated with the increase of the gain $\Lambda/\kappa$ in the DMB  case.  Meanwhile, $\mathcal{S}_{Y_{b_{l}}}$ reaches the maximum value at the optimal phase $\phi=\pi$.  However, the four mechanical modes cannot be squeezed when the OPA is introduced into the system in the DMU case. These results indicate that the OPA provides the physical origin for generating the mechanical squeezing, and the dark-mode breaking is the dominate factor for the existence of the mechanical squeezing when the mechanical modes are connected to heat baths.

\begin{table*}[t]
\centering
\caption{Some reported theoretical proposals and our proposal for the generation of mechanical squeezing. Columns 1, 2, 3, 4, and 5  present the references, the physical mechanisms, the 3dB limit, the squeezing types, and the number of squeezed resonators, respectively.} \label{table1}
\begin{tabular*}{1\textwidth}{@{\extracolsep{\fill}}c c c c c}
\toprule
References                                                                         & Physical mechanisms                      & 3dB limit             & Squeezing  types      &  Number of squeezed resonators  \\ \hline
Agarwal \emph{et al.}~\cite{Agarwal1991}                    &  parametric amplification               &                              &  transient state          &            one                     \\ \hline
Szorkovszky \emph{et al.}~\cite{Szorkovszky2011}     &  quantum measurement                   &  breaking               &  steady state              &           one                      \\ \hline
Mari  \emph{et al.}~\cite{Mari2009}                          & parametric modulation                        &   unbreaking            &  transient state          &          one                       \\ \hline
Liao    \emph{et al.}~\cite{Liao2011}                       & parametric modulation                         &   unbreaking               &  transient state          &         one                 \\  \hline
Kronwald   \emph{et al.}~\cite{Kronwald2013}       & reservoir engineering                            &   breaking              &   steady state            &          one                 \\ \hline
Tan    \emph{et al.}~\cite{Tan2013}                          &  reservoir engineering                            &                              &   steady state           &         two                 \\ \hline
Huang   \emph{et al.}~\cite{Huang2021}                   & reservoir engineering                             &   breaking             &  steady state            &         one                     \\  \hline                                                   Nunnenkamp \emph{et al.}~\cite{Nunnenkamp2010}  & mechanical nonlinearity                      &                               &  steady state            &          one                 \\  \hline
Asjad \emph{et al.}~\cite{Asjad2014}                &  mechanical nonlinearity                                &   breaking             &  steady state           &           one              \\  \hline
L\"{u} \emph{et al.}~\cite{Lu2015a}              &  mechanical nonlinearity                                  & breaking                &  steady state          &          one                 \\  \hline
J\"{a}hne \emph{et al.}~\cite{Jahne2009}       &  squeezed-state transfer                                  &   breaking              &  steady state           &           one                  \\  \hline
Agarwal \emph{et al.}~\cite{Agarwal2016}      &  squeezed-state transfer                                  &   unbreaking          &    steady state         &            one                \\  \hline
This proposal                                                           &  squeezed-state transfer                                &   unbreaking         &   steady state           &        two or multiple           \\  \hline                                                                                            \botrule
\end{tabular*}
\end{table*}

\section{Comparison between the previous theoretical proposals and our proposal\label{comparison}}

Quantum squeezing is an important resource in quantum optics and quantum information science. Currently, there exist some theoretical proposals for generation of mechanical squeezing in optomechanical systems. In this section, we present the comparison between these previous theoretical proposals and our proposal for the generation of mechanical squeezing. In Table~\ref{table1} we present the physical mechanisms for squeezing generations, the 3dB limit, the squeezing types, and the number of  the squeezed resonators for these proposals. We find that many physical mechanisms can be used to generate the mechanical squeezing, including parametric amplification, quantum measurement, parametric modulation, reservoir engineering, mechanical nonlinearity, and squeezed-state transfer. Meanwhile, the types of generated mechanical squeezing include the transient-state squeezing and the steady-state squeezing. Some generated mechanical squeezing breaks the 3dB limit, and some mechanical squeezing does not break the 3dB limit. We should mention that, most of previous schemes considered the generation of quadrature squeezing for a single mechanical mode. However, in the present proposal, we consider the squeezing generations in two and multiple degenerate mechanical modes. In particular, we want to point out that, the main contribution of this proposal is to create mechanical squeezing by controlling the dark-mode effect. To break the dark modes, we introduce the synthetic-gauge-field method. More importantly, we find that this method not only breaks the dark modes and generates the strong steady-state mechanical squeezing, but also improves the tolerance of the mechanical squeezing to thermal noise by about three orders of magnitude. Therefore,  the dark-mode manipulation and the generated single-mode squeezing of multiple mechanical modes are the main results in this proposal.

\section{conclusion \label{conclusion}}

In conclusion, we have used the synthetic-gauge-field method to generate and enhance the quantum squeezing of two mechanical modes in a three-mode optomechanical system consisting of an optical mode and two degenerate mechanical modes. We have found that the  optical squeezing is created by the degenerate OPA, and then it is transferred to the two mechanical modes. In particular, the breaking of the dark-mode effect plays a key role in the appearance of the mechanical squeezing. We have found that this DMB mechanism switches the tolerance of the mechanical squeezing to thermal noise from extremely fragile to extraordinary robust. Especially, the threshold number of thermal phonons for preserving the mechanical squeezing can reach around $100$. Moreover, we have generalized this method to realize the squeezing of multiple mechanical modes by breaking the dark-mode effect. We have also compared our proposal with the previous theoretical proposals for generating the mechanical squeezing. Our results pave a way toward the generation of the macroscopic mechanical squeezing and initiate the study of noise-immune quantum resources.

\begin{acknowledgments}
J.-Q.L. was supported in part by the National Natural Science Foundation of China (Grants No.~12175061, No.~12247105, No.~11822501, No.~11774087, and No.~11935006) and the Science and Technology Innovation Program of Hunan Province (Grants No.~2021RC4029 and No.~2020RC4047). D.-G.L. was supported in part by the Japan Society for the Promotion of Science (JSPS) Postdoctoral Fellowships (Grant No.~P23027).
\end{acknowledgments}


\begin{thebibliography}{99}
\bibitem{Cooling}     We checked the ground-state cooling of two mechanical modes with the synthetic-gauge-field method. These results indicate that the two mechanical modes cannot be cooled into their quantum ground states in the absence of synthetic magnetism ($\eta=0$). However, once the synthetic magnetism is introduced ($\eta>0$), the ground-state cooling of two mechanical modes can be achieved.  Moreover, we found that the better the cooling performance of the two mechanical modes, the greater the degree of squeezing. We also studied the effect of the degenerate OPA (gain $\Lambda$ and phase $\phi$) on the cooling performance of the two mechanical modes in both the DMU and DMB cases. It was found that the DMB is the dominate factor for the ground-state cooling of two mechanical modes. Only when the dark mode is broken, the mechanical modes can be cooled into their quantum ground states.  Meanwhile, the parameters $\Lambda$ and $\phi$ have weak effect on the cooling performance of mechanical modes.
\end{thebibliography}
\end{document}